\documentclass[12pt]{article}
\usepackage{geometry}
\usepackage{amsmath,amssymb}
\usepackage[nosort]{cite}
\usepackage{float}
\usepackage{latexsym}
\usepackage{graphicx}
\usepackage{color}
\usepackage{subfig}
\usepackage{color}

\newcommand{\beq}{\begin{equation}}
\newcommand{\be}{\begin{equation}}
\newcommand{\ee}{\end{equation}}
\newcommand{\bea}{\begin{eqnarray}}
\newcommand{\eea}{\end{eqnarray}}

\newcommand{\pa}{\partial}

\newcommand{\nn}{\nonumber}

\newcommand{\ma}{{\mathcal{A}}}
\newcommand{\mf}{{\mathcal{F}}}
\newcommand{\eA}{e^{A_1}}
\newcommand{\eC}{e^{C_1}}
\newcommand{\eG}{e^{G_1}}
\newcommand{\mc}{{\mathcal{C}}}
\newcommand{\mb}{{\mathcal{B}}}
\newcommand{\rd}[1]{{\color{red} #1}}
\begin{document}

\begin{titlepage}
\hbox to \hsize{\hspace*{0 cm}\hbox{\tt }\hss
    \hbox{\small{\tt }}}

\vspace{1 cm}

\centerline{\bf \Large Black holes in asymptotically Lifshitz spacetimes }

\vspace{.6cm}

\centerline{\bf \Large with arbitrary critical exponent.}

\vspace{1 cm}
 \centerline{\large  $^\dagger\!\!$ Gaetano Bertoldi, $^\dagger\!\!$ Benjamin A. Burrington and $^\dagger\!\!$ Amanda Peet}

\vspace{0.5cm}
\centerline{\it ${}^\dagger$Department of Physics,}
\centerline{\it University of Toronto,}
\centerline{\it Toronto, Ontario, Canada M5S 1A7. }

\vspace{0.3 cm}

\begin{abstract}
Recently, a class of gravitational backgrounds in $3+1$ dimensions have been proposed as holographic duals to a Lifshitz theory describing critical phenomena in $2+1$ dimensions with critical exponent $z\geq 1$.  We numerically explore black holes in these backgrounds for a range of values of $z$.  We find drastically different behavior for $z>2$ and $z<2$.  We find that for $z>2$ ($z<2$) the Lifshitz fixed point is repulsive (attractive) when going to larger radial parameter $r$.  For the repulsive $z>2$ backgrounds, we find a continuous family of black holes satisfying a finite energy condition.  However, for $z<2$ we find that the finite energy condition is more restrictive, and we expect only a discrete set of black hole solutions, unless some unexpected cancellations occur.  For all black holes, we plot temperature $T$ as a function of horizon radius $r_0$.  For $z\lessapprox  1.761$ we find that this curve develops a negative slope for certain values of $r_0$ possibly indicating a thermodynamic instability.
\end{abstract}
\end{titlepage}

\section{Introduction}

Ever since the Maldacena conjecture \cite{Maldacena:1997re}, holography has become an important technique for studying strongly coupled systems (for a review, see \cite{Aharony:1999ti}).  Traditionally, much of the work has been devoted to the study of relativistic field theories in $3+1$ dimensions with superconformal symmetry \cite{Oz:1998hr}.  However, holographic methods have also proven useful in considerably less symmetric situations \cite{smallsym}.  It has also proven useful to construct toy models \cite{toys} imitating the string theory backgrounds, with the surprising result that much of the physics is captured.\footnote{although one prefers string theory ``brane setups,'' where one knows how to describe the weakly coupled degrees of freedom.}

From these $3+1$ dimensional setups, it is known that when describing finite temperature field theories one must consider the black holes in the dual geometry.  The canonical example is that the thermodynamic properties of AdS-Schwarzschild black holes \cite{Hawking:1982dh} match thermal properties of $\mathcal{N}=4$ SYM theories \cite{Witten:1998zw}, exhibiting a host of interesting effects.  The black hole backgrounds themselves have led to interesting conjectures about the nature of strongly coupled plasmas \cite{Gubser:1996de}.  There have also been recent discoveries relating higher derivative corrections in gravitational actions to unitarity in the field theory \cite{causalitycharge}.

Given this, one may be curious about what other types of field theoretic systems can be modeled in terms of a gravitational theory.
Other than the $3+1$ dimensional setups mentioned above, one can also use holography to study $2+1$ dimensional field theories relevant for condensed matter systems.  Recently, much effort has gone into describing quantum critical behavior for these theories using holographic techniques, for a review, see \cite{Hartnoll:2009sz}.  Although not the only type of system one could study, quantum critical systems exhibit a scaling symmetry
\be
t\rightarrow \lambda^{z}t, \quad x_i\rightarrow \lambda x_i
\ee
similar to the scaling invariance of pure AdS ($z=1$) in the Poincar\'e patch.  From a holographic standpoint, this suggests the form of the spacetime metric
\be
ds^2=L^2\left(r^{2z} dt^2+ r^2 dx^i dx^j \delta_{i j}+\frac{dr^2}{r^2}\right), \label{scalemetric}
\ee
where the above scaling is realized as an isometry of the metric along with $r\rightarrow \lambda^{-1} r$.
Other metrics exist with the above scaling symmetry, but also with an added Galilean boost symmetry \cite{Son:2008ye,Balasubramanian:2008dm,Mazzucato:2008tr,Wen:2008hi}.

Often, a good place to begin studying any system is to write down a toy or ``phenomenological'' model \cite{Son:2008ye,Balasubramanian:2008dm} to study generic properties (for thermal versions of these models, see
\cite{Yamada:2008if}).  One may then consider possible embeddings into a more fundamental theory \cite{Mazzucato:2008tr,Yamada:2008if,Azeyanagi:2009pr}, such as a string theory, where more information is known about the weakly coupled physics.  One could also consider gravitational theories which inherently have some nonrelativistic scaling built in \cite{Horava:2009uw}. However, here we will be content to study the relatively simple model considered in \cite{Kachru:2008yh} where Kachru, Liu and Mulligan consider an action which admits a solution with metric (\ref{scalemetric}) (earlier studies of these metrics in a ``brane world'' scenario appear in \cite{Koroteev:2007yp}, and further investigated in \cite{Koroteev:2009xd}).  We leave the possible embedding of this model into string theory for future work, although other kinds of generalizations appear in \cite{Taylor:2008tg,Pang:2009ad,Kovtun:2008qy}.  It should also be noted that a related system \cite{Azeyanagi:2009pr} with anisotropic space scaling has been constructed (in 4+1 dimensions) complete with analytic black brane solutions: this may serve as a template for embedding these types of theories into string theory.

It is the current aim of this work to numerically construct the black hole backgrounds that asymptote to the metric (\ref{scalemetric}) given the equations of motion for the model \cite{Kachru:2008yh}.  Our work can be thought of as complementary to the work of \cite{Danielsson:2008gi}, \cite{Mann:2009yx} where they consider the case $z=2$.

We now turn to the model of \cite{Kachru:2008yh}.  The action that we consider is
\be
S=\int d^4x\sqrt{-g}\left(R-2\Lambda-\frac14 \mf_{\mu \nu} \mf^{\mu \nu}-\frac{c^2}{2}\ma_{\mu}\ma^{\mu}\right)\label{staction}
\ee
where $\mf=d\ma$.
Up to a Legendre transform, this is equivalent to the action given in \cite{Kachru:2008yh}.  This can be seen directly from the equations of motion given in \cite{Kachru:2008yh} with the identification that $*F_3=\frac{1}{c}\ma$ where $F_3$ is the Kachru et. al. three-form, and $\ma$ is our one-form given above.  Further, we will parameterize the constants in the action above as
\be
c=\frac{\sqrt{2{Z}}}{\hat{L}}, \quad \Lambda=-\frac12\frac{Z^2+Z+4}{\hat{L}^2} \label{lifsolcons}.
\ee
Note that this allows for arbitrary negative $\Lambda$, and allows for the ratio of physical constants to be in the regime $\frac{5}{4}\leq \frac{-\Lambda}{c^2} < \infty$.  The specific choice (\ref{lifsolcons}) will become convenient shortly.

There are several known solutions to the above action.  First, we consider the ``black brane'' and ``black hole'' solutions in pure AdS,
\bea
ds^2&=& \left(\frac{-3}{\Lambda}\right)\left(-(\rd{\sigma}+r^2 f(r))dt^2+r^2(dx_1^2+(1-\rd{\sigma \cos^2(x^1)})dx_2^2)+\frac{dr^2}{\rd{\sigma}+r^2f(r)}\right), \nn \\
f(r)&=&1-\frac{r_0(\rd{\sigma}+r_0^2)}{r^3} \nn \\
\ma&=&0.
\eea
Here, and throughout, we will be considering two cases simultaneously: $\sigma=0$ is the ``black brane'' case and $\sigma=1$ is the ``black hole'' solution.  We will color code $\sigma$ and all terms multiplied by this factor in red, so it is easy to see how the various equations are modified (however, $\sigma$ will always appear, so that the color coding is redundant).

There is, of course, also the solution discussed in \cite{Kachru:2008yh}, given by
\bea
ds^2&=&L^2 \left( -r^{2z} dt^2+r^2(dx_1^2+dx_2^2)+\frac{dr^2}{r^2}\right) \nn \\
\ma&=& L^2\frac{r^z}{z}\sqrt{\frac{2z(z-1)}{L^2}}dt \label{lifsol}
\eea
with the identification $z=Z$ and $L=\hat{L}$.  One may of course invert these equations and find $z(\Lambda,c), L(\Lambda,c)$, perhaps more intuitively regarding the parameters of the solution as depending on the parameters of the theory.  However, here we will only be concerned with the restricted set of theories given by (\ref{lifsolcons}).  The above background will serve as the asymptotic form of all our black hole/brane spacetimes as $r \rightarrow \infty$.
\begin{figure}
\centering
\includegraphics[width=0.35\textwidth]{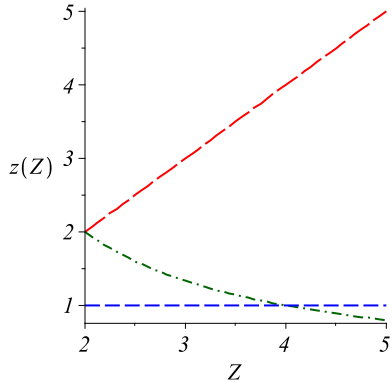}
\caption{The various fixed points for a given action parameterized by $Z$ and $\hat{L}$.  The (dash-dot) green curve is $z=\frac{4}{Z}$, the (long-dashed) red sloped line is $z=Z$ and the (short-dashed) blue flat line is $z=1$.  We choose to plot only $Z>2$ because these specify different actions that have different ratios $\frac{-\Lambda}{c^2}$.  We will find that for the critical exponent $z<2$ (the dash-dot green curve) that the Lifshitz fixed point is ``attractive'' when going to larger $r$ and that for $z>2$ (the red sloped line) it is repulsive.  Further, for $Z>4$ the solution $z=4/Z<1$ is imaginary, and so we exclude it.}
\label{fixedpoints}
\end{figure}

We also note that there may be multiple Lifshitz fixed points given the same action.  The most efficient way to find these is to write down two solutions of the form (\ref{lifsol}) and requiring that even though the values $z,L$ and $z',L'$ are different, the parameters of the action $c^2(Z,\hat{L})|_{Z=z,\hat{L}=L}=c^2(Z,\hat{L})|_{Z=z',\hat{L}=L'}$ and $\Lambda(Z,\hat{L})|_{Z=z,\hat{L}=L}=\Lambda(Z,\hat{L})|_{Z=z',\hat{L}=L'}$ remain unchanged.  A second solution is then found, given by
\be
z'=\frac{4}{Z}, \qquad L'=\frac{2}{Z}\hat{L}.
\ee
Hence, the solution
\bea
ds^2&=&L'^2 \left( -r^{2z'} dt^2+r^2(dx_1^2+dx_2^2)+\frac{dr^2}{r^2}\right) \nn \\
\ma&=& L'^2\frac{r^{z'}}{z'}\sqrt{\frac{2z'(z'-1)}{L'^2}}dt \label{lifsol2}
\eea
is also a solution with the parameters given in (\ref{lifsolcons}).  We can see that $z,z'>1$ for the solutions (\ref{lifsol}) and (\ref{lifsol2}) to be real.  Therefore, there are two solutions for the range $1<Z<4$ because both $Z$ and $4/Z$ are bigger than $1$.  This gives us the following picture: we parameterize the constants of the action as (\ref{lifsolcons}), and so $Z$ and $\hat{L}$ are our free parameters.  There exist solutions of the form (\ref{lifsol}) with the assignment $z=Z, L=\hat{L}$ {\it or} $z=\frac{4}{Z}$ and $L=\frac{2}{Z}\hat{L}$.  The fixed point of this transformation between solutions is $Z=2$, and so to avoid double counting solutions, we consider $Z>2$, which is also the range denoting different actions with different ratios $\frac{-\Lambda}{c^2}$.  Also note that pure $AdS$ is always a solution to the above equations as well (in the global patch when using the sphere) and so $z=1$ with $Z$ unconstrained is also a good fixed point.  Hence, we obtain a plot of fixed points for various $Z$ in figure \ref{fixedpoints}.

The remainder of the paper is dedicated to the analysis of the equations of motion resulting from (\ref{staction}).  We organize this as follows: in the next subsection we write down an Ansatz and the resulting differential equations coming from the action (\ref{staction}).  In section 2, we explore series expansions near a regular horizon ($r=r_0$), and near the asymptotic region, $r=\infty$.  Further, we discuss the finite energy condition, defined as the on shell Hamiltonian for our action \cite{Hawking:1995fd}.  In section 3 we combine the results of section 2, and numerically integrate the equations of motion.  We end section 3 with a discussion of the results, some open questions, and possible future directions.

\subsection{The Ansatz and reduced action.}

As explained above, we will be concerned with constructing the black brane/hole type solutions that asymptote to (\ref{lifsol}).  We do so by considering the Ansatz
\bea
ds^2&=&-\exp(2A(r))dt^2+\exp(2B(r))((dx^1)^2+(1-\rd{\sigma \cos^2(x^1)}) (dx^2)^2) \nn \\
&& \qquad \qquad \qquad \qquad +\exp(2C(r))dr^2 \nn \\
&&\ma=\exp{G(r)}dt.
\eea
Again, in the above, we have introduced two distinct cases: when $\sigma=1$ the two dimensional metric is that of the unit $S^2$ and when $\sigma=0$ it is simply flat two dimensional space.  To obtain the equations of motion, we will first reduce the action to one dimension, plugging in the above Ansatz into action (\ref{staction}).
After integration by parts, this gives a reduced action
\bea
L_{1D}&=&4e^{(2B+A-C)}\pa B \pa A+2e^{(2B+A-C)} (\pa B)^2+\frac12 e^{(-A+2B-C+2G)}(\pa G)^2 \nn \\
&&-2\Lambda e^{(A+2B+C)}+\frac12 c^2 e^{(-A+2B+C+2G)}+ \rd{\sigma 2e^{(A+C)}} \label{actions}
\eea
where we use the abbreviation that $\pa \equiv \frac{\pa}{\pa r}$.  This action reproduces all Einstein-Maxwell equations, if one includes the equation of motion for $C(r)$.  $C(r)$ acts as a Lagrange multiplier imposing the ``zero Hamiltonian'' condition.  The above action contains 3 dynamical fields, and one constraint equation.  We expect, therefore, $6-1=5$ integration constants associated with solutions to the equations near a generic point.  However, we expect to be able to remove one of the remaining constants via coordinate transformations, and so we expect $6-2=4$ constants.  Using coordinate transformations to fix some relation we will refer to as gauge fixing.

The above action gives the following equations of motion
\bea
4\pa(e^{(A+2B-C)}(\pa A))+2e^{(A+2B-C)}(\pa A)^2-2e^{(A+2B-C)}(\pa A+ \pa B)^2&&\nn \\
\kern-2em -\frac32 e^{(-A+2B-C+2G)} (\pa G)^2-\frac32 c^2 e^{(-A+2B+C+2G)}+2\Lambda e^{(A+2B+C)}+{\color{red}{\sigma2e^{(A+C)}}}&=&0 \label{eqas} \\
\kern-2.5em -4\pa(e^{(A+2B-C)}(\pa A+\pa B)) -4e^{(A+2B-C)} (\pa A)^2 +4e^{(A+2B-C)}(\pa A+ \pa B)^2&& \nn \\
 e^{(-A+2B-C+2G)}(\pa G)^2 +e^{(-A+2B+C+2G)}c^2-4\Lambda e^{(A+2B+C)}&=&0  \label{eqabs}\\
-\pa(e^{(-A+2B-C)}\pa e^{(G)})+c^2 e^{(-A+2B+C)}e^{(G)}&=&0 \label{eqgs}\\
-4e^{(2B+A-C)}\pa B \pa A-2e^{(2B+A-C)} (\pa B)^2-\frac12 e^{(-A+2B-C+2G)}(\pa G)^2 && \nn \\
-2\Lambda e^{(A+2B+C)}+\frac12 c^2 e^{(-A+2B+C+2G)}+{\color{red}{\sigma2e^{(A+C)}}}&=&0. \label{Hams}
\eea
where the last equation is the Hamiltonian constraint.  We will be concerned with turning on a ``blackening'' factor for those backgrounds found in \cite{Kachru:2008yh}; these backgrounds all have $c\neq0$.  However, in the above we can see why the $c=0$ (massless) case is special.  In such a case, one can shift $A\rightarrow A+\delta$ and $C\rightarrow C-\delta$ with $\delta$ being an arbitrary function of $r$.  Doing so leaves the equation of motion for $G$ given by (\ref{eqgs}) unaffected (the Laplacian part remains unchanged).  In such a situation, $G$ may be left alone, and still satisfy its equation of motion. The mass parameter $c$ changes this because of the relative minus sign in front of $C(r)$ in the exponential.  Hence, we expect the massive vector field above to change in a ``blackened'' background.

Before trying to find solutions to the above equations, we present a first integral for the $\sigma=0$ version of the above equations of motion.  One may infer the existence of this first integral because $\sigma=0$ leaves only 2 potential terms in the effective Lagrangian.  However, because there are 3 fields some linear combination of them does not couple to the potential.  One may also see this as the Noether charge associated with the shift
\be
\begin{pmatrix}
A(r) \\
B(r) \\
C(r) \\
G(r) \\
\end{pmatrix}
\rightarrow
\begin{pmatrix}
A(r)+ \delta \\
B(r)- \frac{\delta}{2}\\
C(r)+ 0 \\
G(r)+ \delta  \\
\end{pmatrix}
\ee
with $\delta$ a constant.  The above represents a diffeomorphism which preserves the volume element $dt dx^1 dx^2$.  This is why it is inherited as a Noether symmetry in the reduced Lagrangian.

Rather than writing the conserved quantity, we write out the differential equation that may be directly integrated in the $\sigma=0$ case:
\be
\pa(2e^{(A+2B-C)}\pa A- 2e^{(A+2B-C)}\pa B-e^{(-A+2B-C+2G)} \pa G)+\rd{\sigma 2e^{(A+C)}}=0 \label{D0}.
\ee
As an example, (for $\sigma=0$) we integrate once, and plug in the black brane in AdS case to find
\be
2e^{(A+2B-C)}\pa A- 2e^{(A+2B-C)}\pa B-e^{(-A+2B-C+2G)} \pa G \equiv D_0=3\frac{-3}{\Lambda}r_0^2
\ee
so the above equation is indeed satisfied.

Given the above considerations, we may pick as a complete set of differential equations (\ref{Hams}), (\ref{D0}) and one of the three second order differential equations (\ref{eqas}) or (\ref{eqabs}) or (\ref{eqgs}).  In fact, one can gauge fix one field to a known function (e.g. $e^{B}=r$).  This will transform one of the three second order differential equations into a first order differential equation.  In the $\sigma=0$ case, one is therefore left solving 3 first order differential equations (after integrating (\ref{D0}) once).  While this seems suggestive, we have been unable to use this fact to solve the equations exactly.

In general, we will want to explore solutions to the equations of motion that asymptote correctly to (\ref{lifsol}).  For this reason, we take the following definitions
\bea
&&A(r)=\ln(r^z L)+A_1(r), \qquad B(r)=\ln(rL)+B_1(r) \nn \\
&&C(r)= \ln\left(\frac{L}{r}\right)+C_1(r), \qquad G(r)=\ln\left(\frac{L^2r^z}{z}\sqrt{\frac{2z(z-1)}{L^2}}\right) +G_1(r)
\eea
As mentioned above, the 3 second order differential equations may be reduced to two first order and one second order differential equations by a gauge choice, and the presence of a conserved Hamiltonian.  In the black brane ($\sigma=0$) case, one can further reduce to 3 first order equations due to the scaling symmetry, however we will deal with the $\sigma=1$ case here.  While we could make many different gauge choices, what will be important for us later is the gauge choice
\be
B_1(r)=0.
\ee
In such a gauge choice, the second order differential equation for $B$ becomes a first order equation.  In addition, we take the Hamiltonian, and the second order equation for $e^G$.  Using the first two of these three equations, we may eliminate $\pa e^{A_1}$ and $\pa e^{C_1}$ in the third, which we do.  Further, we isolate $\pa e^{A_1}$ and $\pa e^{C_1}$ in their own equations.  Doing so, we find the three equations
\bea
&&\kern-1.5em\pa \eA + \frac14 \frac{r\left(\pa \eG \right)^2(z-1)}{z \eA}+\frac12 \frac{\eG \pa \eG(z-1)}{\eA} \nn \\
&&\kern-1.5em \quad +\frac14 \frac{2ze^{2A_1}(1+2z)-ze^{2A_1}e^{2C_1}(z^2+z+4)+ z\left(z-2e^{2C_1}\right)e^{2G_1}(z-1)}{z r \eA }\nn \\
&&\qquad \qquad \qquad \qquad \qquad \qquad \qquad \qquad \qquad-{\color{red}\sigma\frac12 \frac{\eA e^{2C_1}}{r^3}}=0 \label{A1eq}\\
&&\kern-1.5em\pa \eC - \frac14 \frac{r\eC\left(\pa \eG \right)^2(z-1)}{z e^{2A_1}}-\frac12 \frac{\eC \eG \pa \eG(z-1)}{e^{2A_1}} \nn \\
&&\kern-1.5em \quad -\frac14 \frac{\eC\left(6ze^{2A_1}-ze^{2A_1}e^{2C_1}(z^2+z+4)+ z\left(z+2e^{2C_1}\right)e^{2G_1}(z-1)\right)}{z r e^{2A_1} }\nn \\
&&\qquad \qquad \qquad \qquad \qquad \qquad \qquad \qquad \qquad+{\color{red}\sigma\frac12 \frac{e^{3C_1}}{r^3}}=0 \label{C1eq}\\
&&\kern-1.5em(z-1) \pa^2 \eG + \frac{(z-1)\pa \eG \left(2(z+1)e^{2A_1}-e^{2C_1}e^{2G_1}(z-1)\right)}{e^{2A_1}r} \nn \\
&&\qquad \frac{z(z-1)\eG\left(-e^{2C_1}e^{2G_1}(z-1)+e^{2A_1}(z+1)-2e^{2C_1}e^{2A_1}\right)}{e^{2A_1}r^2}=0 \label{G1eq}
\eea
where we have explicitly left the $z-1$ multiplying the second order equation, to stress that this equation does not need to be solved in the $z=1$ case.  Further, one may combine the first two lines into the following equation
\bea
r\pa\left(\eA \eC\right)+\eC\eA\left(1-\left(\frac{\eC \eG}{\eA}\right)^2\right)(z-1)=0. \label{eCeAeq}
\eea
This merely serves as a check: in the case $z=1$ the black brane/black hole solutions have $\eA \eC=1$, which solves the above equation.  We will use (\ref{eCeAeq}) to find an exact solution for $z=4$ in appendix \ref{exactsolution}.

The initial conditions needed to solve the above equations numerically are initial values for $\eA, \eC, \eG$ and an initial value for $\pa \eG$.

\section{Analytic explorations.}

\subsection{The perturbed solution near the horizon}

We begin by first exploring the solution near the horizon.  We require that $e^{2A}$ goes to zero linearly, $e^{2C}$ has a simple pole, and $e^{G}$ goes to zero linearly to make the flux $d\ma$ go to a constant (in a local frame or not).  Further, we take the gauge $B(r)=\ln(Lr)$ for this section.  We expand
\bea
&&A(r)=\ln\left(r^z L\left(a_0(r-r_0)^{\frac12}+a_0a_1(r-r_0)^{\frac32}+\cdots\right)\right), \qquad B(r)=\ln(rL) \nn \\
&&C(r)= \ln\left(\frac{L}{r}\left(c_0(r-r_0)^{-\frac12}+c_1(r-r_0)^{\frac12}+\cdots\right)\right), \\ &&G(r)=\ln\left(\frac{L^2r^z}{z}\sqrt{\frac{2z(z-1)}{L^2}}\left(a_0g_0(r-r_0)+a_0g_1(r-r_0)^2+\cdots\right)\right). \nn
\eea
Note that by scaling time we can adjust the constant $a_0$ by an overall multiplicative factor (note the use of $a_0$ in the expansion of $G(r)$ as well, as $e^{G}$ multiplies $dt$ for the one-form $\ma$).  We will need to use this to fix the asymptotic value of $A(r)$ to be exactly $\ln(r^zL)$ with no multiplicative factor inside the log.

We plug this expansion into the equations of motion arising from (\ref{actions}), and solve for the various coefficients.  We find a constraint on the 0th order constants: as expected not all boundary conditions are allowed.  We solve for $c_0$ in terms of the other $g_0$ and $r_0$, and find \footnote{We may compare this to the result quoted in \cite{Danielsson:2008gi} and \cite{Mann:2009yx} by taking
$g_0=h_0 c_0 \frac{2}{L}\frac{z}{r_0}\sqrt{\frac{L^2}{2z(z+1)}}$
and resolving for $c_0$.  This gives
$c_0=\frac{r_0^{\frac32}}{\sqrt{{\color{red}\sigma}+r_0^2\left(\frac{z^2+z+4}{2}-h_0^2\right)}}$.  In agreement with their results after identifying our ``$c_0$'' is their ``$g_0$.''}
\bea
c_0 = \frac{\sqrt{(2z+g_0^2 r_0(z-1))} r_0^{\frac32}}{\sqrt{z}\sqrt{{{\color{red}2\sigma }+(z^2+z+4)r_0^2}}}. \label{c0equ}
\eea
The higher order coefficients are
\bea
a_1=\frac{r_0^4(z-1)^2 g_0^4+2r_0(z-1)(r_0^2(z^2+2z+4)+{\color{red}\sigma})g_0^2-4z(r_0^2z(z^2+z+4)+{\color{red}\sigma(2z+1)})} {4r_0z({\color{red}2\sigma}+r_0^2(z^2+z+4))}\nn \\
\; \\
c_1=c_0\frac{3r_0^4(z-1)^2g_0^4-2r_0(z-1)(r_0^2(z^2-2z+4)+{\color{red}\sigma})g_0^2 +4z(r_0^2(z^2+z+4)+{\color{red}3\sigma})}{ 4r_0z({\color{red}2\sigma}+r_0^2(z^2+z+4))} \nn \\
\; \\
g_1=g_0\frac{r_0^4(z-1)^2g_0^4+4zr_0^3(z-1)g_0^2-2z(r_0^2(z^3+2z^2+3z+4)+{\color{red}2\sigma(z+1)})} {2r_0z({\color{red}2\sigma}+r_0^2(z^2+z+4))}. \nn \\
\;
\eea
Again, in the above expressions one must simply drop the terms highlighted in red (or $\sigma=0$) to get the boundary conditions for the ``black brane'' case, removing the term arising from the $S^2$.  One can easily see that this corresponds to a large $r_0$ limit, a feature shared with the black hole/black brane solutions in $AdS_4$.

\subsection{The perturbed solution near r=infinity}
\label{rinftysect}
We now turn to the question of the deformation space around the solution given in (\ref{lifsol}) and (\ref{lifsolcons}).  For this, we will need to consider the term in the potential $\sigma e^{(A+C)}$ to be first order in $\epsilon$ (the perturbative parameter) already.  This is because the ``background solution'' is only a solution to the action (\ref{actions}) with $\sigma=0$.  In fact, one can see this from the equations of motion written as (\ref{A1eq})-(\ref{G1eq}), where the terms multiplied by $\sigma$ have more powers of $r$ in the denominator, and so may be neglected in the large $r$ limit.  Therefore, we take the expansion of the functions
\bea
&&A(r)=\ln(r^z L)+\epsilon A_1(r), \qquad B(r)=\ln(rL)+\epsilon B_1(r) \nn \\
&&C(r)= \ln\left(\frac{L}{r}\right)+\epsilon C_1(r), \qquad G(r)=\ln\left(\frac{L^2r^z}{z}\sqrt{\frac{2z(z-1)}{L^2}}\right) +\epsilon G_1(r)
\eea
and regard the term $\sigma e^{(A+C)}$ to be a small correction to the action to find the equations of motion
\bea
&&\kern-4em-\pa^2 A_1 + \frac{1}{r}\pa\left(-2(z+1)A_1-(z-1)B_1+\frac32(z-1)G_1+C_1\right)  \nn \\
&&\kern-3em+\frac{1}{r^2}\left(-\frac32 \left(z^2+z-2\right)A_1+\frac32\left(z^2+z-2\right)G_1
+\frac12\left(z^2+7z-2\right)C_1\right) \nn \\
&&\kern20em-\rd{\sigma\frac{1}{2r^4}}=0 \label{eoma1}\\
&&\kern-4em-\pa^2 (B_1)+\frac{1}{r}\pa \left(-4B_1-\frac12(z-1)G_1+C_1\right)  \nn \\
&&\kern-3em+\frac{1}{r^2}\left(\frac12(z^2+z-2)A_1-\frac12(z^2+z-2)G_1+\frac12(z^2+z-6)C_1\right) \nn \\
&&\kern20em+\rd{\sigma\frac{1}{2r^4}}=0  \label{eomab1}\\
&&\kern-4em-\pa^2G_1 + \frac{1}{r}\pa \left(z A_1-2 B_1 -(z+3) G_1 +z C_1\right)+\frac{4z C_1}{r^2}=0.\label{eomg1}
\eea

In addition, we have the Hamiltonian constraint
\bea
 &&\kern-1em \frac{1}{r}\pa\left(-2 A_1-2(z+1)B_1-(z-1)G_1\right) \nn \\
&&\kern-1em \quad +\frac{1}{r^2}\left((z^2-3z+2)A_1-(z^2-3z+2)G_1+(z^2+3z+2)C_1\right)+\rd{\sigma\frac{1}{r^4}}=0.\label{ham1}
\eea
Note that in the above expressions we have not yet taken the $B_1(r)=0$ gauge.  We turn to this choice shortly.

We may read the above equations in the following way:  The terms with fields in them are in fact the perturbed equations associated with action (\ref{actions}) with $\rd{\sigma}=0$, and the source terms come from plugging in the 0th order solution into the correction term in the action $\rd{\sigma e^{(A+C)}}$.  These terms are easily identifiable as those being non homogeneous in powers of $r$.

Using straightforward perturbation theory, we may find the solutions in the $B_1=0$ gauge
\bea
A_1(r)&=& {\mathcal{C}}_0 \frac{(z-1)(z-2)}{(z+2)} r^{-z-2}
+\mc_1\left(z^2+3z+2+(z+1)\gamma\right)r^{-\frac{z}{2}-1+\frac{\gamma}{2}}  \label{A1expansion} \\
&&+\mc_2\left(z^2+3z+2-(z+1)\gamma\right)r^{-\frac{z}{2}-1-\frac{\gamma}{2}}
-z \mb_0+\rd{\sigma\frac{1}{2r^2(z^2-2z+2)}} \nn \\
B_1(r)&=&0  \label{B1expansion}\\
C_1(r)&=& -{\mathcal{C}}_0(z-1)r^{-z-2}+ \mc_1\left(z^2-7z+6+(z-1)\gamma\right)r^{-\frac{z}{2}-1+\frac{\gamma}{2}} \label{C1expansion} \\
&& +\mc_2\left(z^2-7z+6-(z-1)\gamma\right)r^{-\frac{z}{2}-1-\frac{\gamma}{2}}-\rd{\sigma\frac{1}{2r^2(z^2-2z+2)}} \nn \\
G_1(r)&=& {\mathcal{C}}_0 \frac{2(z^2+2)}{z+2}r^{-z-2}+\mc_1 4z(z+1)r^{-\frac{z}{2}-1+\frac{\gamma}{2}} \label{G1expansion} \\
&& +\mc_24z(z+1)r^{-\frac{z}{2}-1-\frac{\gamma}{2}} -z \mb_0+\rd{\sigma\frac{1}{r^2(z^2-2z+2)}} \nn
\eea
where we have defined the useful constant
\be
\gamma=\sqrt{9z^2-20z+20}.
\ee
We postpone plotting the exponents of the various powers until section \ref{conclusions} where we will discuss how the finite energy condition constrains the constants $\mc_0,\mc_1,\mc_2,\mb_0$.  The graphs appear in figure \ref{alphapowers} of that section.

The rescaling of time is manifested by shifting $\mb_0$, but no such constant seems to exist for rescaling $x_i$ (for the $\rd{\sigma}=0$ case).  These are in fact gauge equivalent given the symmetry of the background $r\rightarrow \lambda r, t\rightarrow \lambda^{-z}t, x_{1,2}\rightarrow \lambda^{-1} x_{1,2}$, see appendix \ref{gaugesection}.

Above, one must read the powers $r^{-\frac{z}{2}-1-\frac{\gamma}{2}}$ and $r^{-1-\frac{z}{2}+\frac{\gamma}{2}}$ carefully.  This is because for certain values of $z$ these powers become the same as other modes already appearing in the expansion.  For example, in the limit that $z=2$ the mode $r^{-1-\frac{z}{2}+\frac{\gamma}{2}}$ becomes a constant.  This constant interferes with the $\mb_0$ constant term.  To get the required number of independent integration constants, one must take $\mb_0, \mc_1$ to have $z$ dependent pieces: the leading orders as $z\rightarrow 2$ are engineered to cancel leaving the next term in the expansion of $\lim_{z\rightarrow 2} r^{-\frac{z}{2}-1-\frac{\gamma}{2}}$ (this is a $\ln(r)$ piece).  Doing so consistently will leave the term $r^{-1-\frac{z}{2}+\frac{\gamma}{2}}$ in $C_1(r)$ to become a constant, even though it naively goes to zero when $z=2$.  Similar considerations appear for the power $r^{-\frac{z}{2}-1-\frac{\gamma}{2}}$.  In fact, for the $z=2$ case, we get modes of the form $\mb_0$, $\ln(r)$, $r^{-4}$ and $\ln(r)r^{-4}$ using the above considerations \cite{Danielsson:2008gi}.

\subsection{Finite energy condition.}
\label{finiteensect}

Here we explore the finite energy conditions for our backgrounds.  We follow the discussion in \cite{Hawking:1995fd} which in turn follows the conventions of \cite{Wald:1984rg}.  The on-shell Hamiltonian reduces to a boundary integral defined by the intersection of surfaces of constant ``time'' and ``spatial infinity.''  For this, one needs to have a (monotonic) time coordinate, and for us this is simply $t$.  Further, we need to define a ``spatial infinity'' at a given time slice $t$; here again, we simply use ``$r=$ large constant'' to define slicing the spacetime near spatial infinity.  Further, in the black brane case, one must also restrict the integration over $x_1$ and $x_2$, and so we must define an energy density per unit two volume, rather than an absolute energy.  Therefore, we will also need the constant $x^i$ slices as well, and their associated normal vectors.

\begin{figure}
\begin{center}
\includegraphics[width=0.35\textwidth]{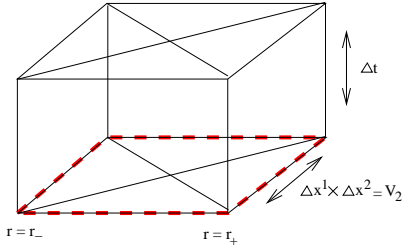}
\end{center}
\caption{The above gives a picture of the cell of spacetime under consideration.  $r_{-}$ is the interior surface, $r_{+}$ is the outer surface with $r_+>r_-$.  The red (dashed) line gives the surface integral used to define the energy given at the bottom time slice.  $r_{-}$ will be taken to approach the horizon (or 0), and $r_+$ will be taken to go to infinity, and further $\Delta x^i$ are both held fixed, with their product $\Delta x^1\times \Delta x^2=V_2$.}
\label{finiteenpic}
\end{figure}

Recall, the metric in the black brane case is
\be
ds^2=-e^{2A(r)}dt^2+e^{2B(r)}\left(\left(dx^1\right)^2+\left(dx^2\right)^2\right)+e^{2C(r)}dr^2
\ee
and so we define the normal vectors to constant $t,r,x^i$ slices as
\bea
n_t^\mu\pa_\mu&=&e^{-A(r)}\pa_t \nn \\
n_r^\mu\pa_\mu&=&e^{-C(r)}\pa_r  \\
n_i^\mu\pa_\mu&=&e^{-B(r)}\pa_{x^i} \nn
\eea
Next, we define $\hat{t}^\mu\pa_\mu t=1,\hat{r}^\mu\pa_\mu r=1,\left(\hat{x}^i\right)^\mu\pa_\mu x^j=\delta^{ij}$.  This then allows us to define a lapse function and shift vector for each of these slicings as $\hat{t}^\mu=(N_t n_t^\mu+N_t^\mu)$ and similarly for the other vectors.  For us, the shift vectors are trivial ($N_t^\mu=N_r^\mu=N_i^\mu=0$), and the lapse functions $N_t=e^{A(r)},N_r=e^{C(r)},N_i=e^{B(r)}$.  These are needed to define the measure of the various integrals.  We will consider doing a finite integral and taking the limits of $r_{-}\rightarrow 0$ (or to the horizon) and $r_{+}\rightarrow \infty$.  In figure \ref{finiteenpic} we give a pictorial representation of the relevant integrals.
In \cite{Hawking:1995fd}, it was found that the integrated hamiltonian density on a time slice was given by\footnote{In our normalization, we have a coefficient ``$1$'' in front of the $R$ term in the action.  Therefore, to arrive at our normalization, we simply replace the terms $16 \pi \rightarrow 1, 8\pi \rightarrow \frac12$ in their expressions.}
\be
-2\int_{\Sigma_{t,r^i}} \left(N_t \,^2\!K-N_t^\mu p_{\mu \nu} \hat{r_i}^\mu\right)
\ee
where we have used a generic $\hat{r_i}$ to denote any of the spatial unit vectors (and $r^i$ to denote which boundary we are talking about: either $r=$ constant, or $x^i=$ constant), and $p_{\mu \nu}$ is the momentum conjugate to the time derivative of the metric on the time slice.    For us, this second term is not present because all shift vectors $N_{\{t,r,1,2\}}^{\mu}=0$. The term $\,^2\!K$ is the extrinsic curvature of the spatial boundary slice {\it in} the constant time slice $t={\rm constant}$.
We therefore have 6 integrals to perform, for the 2 limiting values of $(r,x^1,x^2)$.  This gives
\bea
E&=& -2\int dx_i^2 \left[N_t  \left(g^{\mu \nu}+n_t^\mu n_t^{\nu}-n_r^\mu n_r^\nu\right)\nabla_\mu n_{r \nu}\right]|^{r=r{+}}_{r=r_-} \nn \\
&& -2\int dx_2 dr \left[N_t  \left(g^{\mu \nu}+n_t^\mu n_t^{\nu}-n_1^\mu n_1^\nu\right)\nabla_\mu n_{1 \nu}\right]|^{x^1=x^{1+}}_{x^1=x^{1-}} \label{enints} \\
&&-2\int dx_1 dr \left[N_t  \left(g^{\mu \nu}+n_t^\mu n_t^{\nu}-n_2^\mu n_2^\nu\right)\nabla_\mu n_{2 \nu}\right]|^{x^2=x^{2+}}_{x^2=x^{2-}}\nn
\eea
where the bounds of any $x^i$ integration are $(x^{i-},x^{i+}$ and that of any $r$ integration are $(r^-,r^+)$.
The last two terms above are zero because the integrand is independent of $x^1$ and $x^2$.  Further, the first integrand is independent of $x^1$ and $x^2$ and so the integral is just $V_2$.  Recall that $x^i$ are unitless, so at the end we will need to take some factors of $L$ along with this unitless $V_2$ to honestly get an energy density.  This then simplifies to
\bea
\rho &=&\frac{E}{V_2}=-2\left[e^{A(r)}e^{2B(r)}e^{-2B(r)}\delta^{ij}\nabla_i n_{(r)j}\right]|^{r=r{+}}_{r=r_-}\nn \\
&=&2\left[e^{A(r)}\delta^{ij}\Gamma^{r}_{ij} e^{C(r)}\right]|^{r=r{+}}_{r=r_-} \nn \\
&=&-2\left[e^{A(r)}e^{-C(r)}\left(\pa_r\left(e^{2B(r)}\right)\right)\right]|^{r=r{+}}_{r=r_-}. \label{encalc1}
\eea
In the above equation the combination $e^{-C}\pa_r$ is r-diffeomorphism invariant, and so the above is well defined for any redefinition of the $r$ coordinate.  In the pure Lifshitz background the limit $r_-\rightarrow 0$ gives a zero answer.  Further, we note that for the black hole backgrounds we are dealing with, $e^{A}\rightarrow 0$, $e^{-C(r)}\rightarrow 0$, and $e^{B(r)}\rightarrow$ a constant when $r\rightarrow r_0$.  This means that in both cases, when one takes $r_-$ to its limiting value, this boundary will not contribute.  Therefore, only the boundary at $r=r_+\rightarrow \infty $ will contribute.  To render this finite, we will use a background subtraction technique.  Keeping in mind that the zeroth order Lifshitz solution is what we compare to, and working in the $B_1=0$ gauge, we will find that the $r^{-z-2}$ mode is the mode that has finite energy.  Any function that does not fall off at least this fast will give an infinite energy density contribution, even if the background asymptotes to the Lifshitz solution. We will discuss this further in the next section.

One may repeat the above calculation for $\sigma=1$.  One need not consider taking finite two volume in this case because the $S^2$ is compact.  Therefore, the analogs of the last four (out of six) terms of (\ref{enints}) are not present.  Further, in equation (\ref{encalc1}) one replaces $\delta^{ij}\rightarrow g_{S^2}^{ij}$.  However, it is still true that $\Gamma^r_{ij}=-\frac{1}{2} \pa\left(e^{B(r)}\right) g_{(S^2)ij}$.  Given this, the same formula above applies, with the simple replacement $V_2\rightarrow 4\pi$ to account for the finite volume of the unit $S^2$.

Again, we use the $B(r)=\ln(Lr), B_1(r)=0$ gauge.  In such a situation, the above energy reads
\be
\rho=\lim_{r_+\rightarrow \infty} -4\left[e^{A(r)}e^{-C(r)}r\right]|_{r=r{+}}
\ee
for all backgrounds.

\section{Numeric integration, results, and discussion.}
\label{conclusions}

First, we will consider the finite energy condition applied to the large $r$ region of our backgrounds.  We subtract a reference background to obtain a relative energy density
\be
\rho_{\rm rel}=\lim_{r_+\rightarrow \infty}-4\left[r\left(e^{A(r)}e^{-C(r)}-e^{A_{\rm ref}(r)}e^{-C_{\rm ref}(r)}\right)\right]|_{r=r{+}}.
\ee
We require both backgrounds to be asymptotic to the Lifshitz fixed point.  Therefore, we expect that for $r\rightarrow \infty$ the perturbation theory developed in section \ref{rinftysect} applies.  This gives
\bea
\rho_{\rm rel}&&=\lim_{r_+\rightarrow \infty}-4\left[r\left(e^{A(r)}e^{-C(r)}-e^{A_{\rm ref}(r)}e^{-C_{\rm ref}(r)}\right)\right]|_{r=r{+}} \nn \\
&&\approx \lim_{r \rightarrow \infty}-4\left[r^{z+2}\left(A_1(r)-C_1(r)-\left(A_{1,{\rm ref}}(r)-C_{1,{\rm ref}}(r)\right)\right)\right]. \label{finiteenexp}
\eea
\begin{figure}[ht]
\centering
\includegraphics[width=0.45\textwidth]{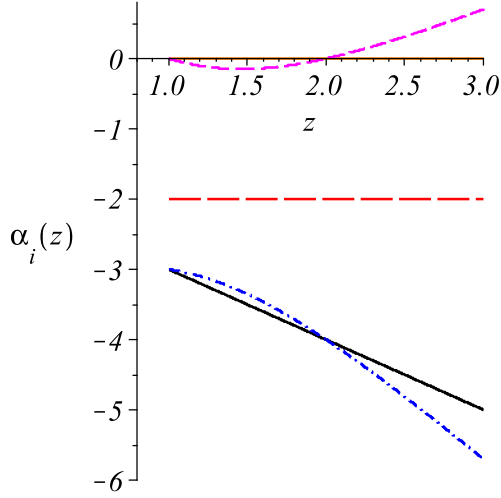}
\caption{Above we graph the exponents of the modes at infinity $r^{\alpha_i(z)}$ as a function of z.  The (solid) black line ($-z-2$) represents the proper ``energy'' mode.  The (long-dashed) red line $(-2)$ represents the inhomogeneous mode induced by the $S^2$.  The line $(-2)$, and all its non-linear descendants, will be universal for backgrounds with the sphere.  All other curves that are above the black line $(-z-2)$ represent infinite energy modes, and curves below this line represent modes with $0$ contribution to the energy of the background.  The top (short-dashed) magenta curve is $\left(\frac{-z}{2}-1+\frac{\gamma(z)}{2}\right)$ (recall this mode has coefficient $\mc_1$), which always represents an infinite energy mode.  The bottom (dash-dot) blue curve is $\left(\frac{-z}{2}-1-\frac{\gamma(z)}{2}\right)$ (recall this mode has coefficient $\mc_2$).  This curve represents a finite energy mode for $z>2$ and an infinite energy mode for $1\leq z\leq 2$: the value $z=2$ gives the curve where this curve crosses the line $-z-2$.}
\label{alphapowers}
\end{figure}

We plot the exponents $r^{\alpha(z)}$ in the modes of (\ref{A1expansion})-(\ref{G1expansion}) in figure \ref{alphapowers}.  We first discuss the $\sigma=0$ case, where the $r^{-2}$ term is not present, and in fact we may use the regular Lifshitz background as our reference ($A_{1,{\rm ref}}(r)=C_{1,{\rm ref}}(r)=0$).  In such a case we require that the coefficient of $r^{-\frac{z}{2}-1+\frac{\gamma}{2}}$ be zero, $\mc_1=0$, so that the limit (\ref{finiteenexp}) does not diverge.  From figure \ref{alphapowers}, we see that after setting $\mc_1=0$ for $z>2$, the next to leading mode at infinity is the $r^{-z-2}$ mode.  This precisely cancels the $r^{z+2}$ term in (\ref{finiteenexp}), so that the $r\rightarrow \infty$ limit is finite.  Hence, for $z>2$ we need to cancel one term at infinity: we must set $\mc_1=0$.  Now recall that we have two parameters at the horizon, and therefore we expect a one parameter family of solutions with finite energy.  This single parameter is most naturally taken to be the position of the horizon.

However, in the case that $(1\leq z \leq 2)$, \footnote{we include the $=$ in these bounds because we get logarithmic terms at the end points: these still have logarithmically divergent energy density} there is an additional mode we must remove, namely $r^{-\frac{z}{2}-1-\frac{\gamma}{2}}$.  This gives two conditions at infinity.  Recall that we always have two constants at the location of the horizon.  Hence, on general grounds we expect that there is at most a discrete set of solutions that have a regular horizon and have finite energy density in the regime ($1\leq z \leq 2$).  However, this is actually not the case for $\sigma=0$.

For the $\sigma=0$ black branes, the scaling symmetry $t\rightarrow \lambda^z t, x^i\rightarrow \lambda x^i, r\rightarrow \lambda^{-1}r$ is exact for the background: only the presence of the horizon breaks this.  This implies that in the $\sigma=0$ case, if there exists one finite energy black brane (i.e. one finds one combination of $g_0$ and $r_0$ that gives a finite energy density black brane), with horizon position $r_0$, then a continuous one parameter family of solutions must exist.  This is because we may apply the scaling symmetry and find a new finite energy black brane, continuously scaling the location of the horizon $r_0\rightarrow \lambda r_0$.  One can see this most easily by considering the correction functions in the metric and form field.  Assume we have a finite energy density solution
\bea
ds^2&=&L^2 \left( -r^{2z} e^{2A_1(g_0,r_0,r)}dt^2+r^2(dx_1^2+dx_2^2)+e^{2C_1(g_0,r_0,r)}\frac{dr^2}{r^2}\right) \nn \\
\ma&=& L^2\frac{r^z}{z}\sqrt{\frac{2z(z-1)}{L^2}}e^{2G_1(g_0,r_0,r)}dt.
\eea
After performing the rescaling, the solution reads
\bea
ds^2&=&L^2 \left( -r^{2z} e^{2A_1(g_0,r_0,\lambda^{-1}r)}dt^2+r^2(dx_1^2+dx_2^2)+e^{2C_1(g_0,r_0,\lambda^{-1}r)}\frac{dr^2}{r^2}\right) \nn \\
\ma&=& L^2\frac{r^z}{z}\sqrt{\frac{2z(z-1)}{L^2}}e^{2G_1(g_0,r_0,\lambda^{-1}r)}dt.
\eea
Now, if $e^{2A_1(g_0,r_0,r)}$ has a linear zero at $r=r_0$, then $e^{2A_1(g_0,r_0,\lambda^{-1}r)}$ has a linear zero at $r=\lambda r_0$ (and so forth for the other functions).  Further, no new terms in the expansion at infinity have been introduced.  Hence, the scaled solution is also a finite energy density black brane with respect to the uncorrected Lifshitz background.

Thus, this scaling symmetry allows us to conclude the following result for the black brane case: either there is a 1 parameter class of black branes (i.e. $\mc_1=0$ implies $\mc_2=0$ for physical input on the horizon $g_0,r_0$), or there are none at all.  It would be interesting to see if there are any by shooting in from infinity and seeing if a regular horizon develops, given a finite energy perturbation at infinity.  If one does not see a regular horizon, then one may conclude that there are {\it no} finite energy density black brane solutions, and conversely if there is one, then there is a continuous family.  We will see later that for the $z>2$ case a continuous class of black branes exist.  However, applying the above scaling argument to the $1<z<2$ case, we see that the naive expectation that there are a discrete set of black branes is incorrect; actually there are either none, or a continuous class of them.

At exactly $z=1$, the counting of constants at infinity is different because one-form field is set to zero.  In such a situation one may count $2$ dynamical functions.  As we saw in the perturbative expansion, the Hamiltonian constraint and gauge fixing remove two additional constants, leaving two constants at infinity.  One of these must be associated with time diffeomorphisms, and so we arrive at only having one meaningful constant at infinity.  Setting $z= 1$, we also only have one constant at the location of the horizon.  Therefore, if the mode found at infinity is finite energy, then one expects a one parameter family of solutions.  If one wishes to turn on the one-form field, one would instead count 2 independent modes at the horizon and 3 at infinity.  This counting is of course valid for both $\sigma=1$ or $\sigma=0$.

Much of the above discussion is directly applicable to the $\sigma=1$ case.  For $z>2$ we again see that we must cancel the top mode $\mc_1=0$.  We also have the $r^{-2}$ term, and all of its higher order corrections to deal with.  However, the $r^{-2}$ comes with a known coefficient, and so do all of its ``descendent'' modes.  So, we get $r^{-4},r^{-6},$ etc. terms, all with known coefficients, which are the same for all solutions.  Therefore, we expect these modes cancel against any given reference $\sigma=1$ background that asymptotes to the Lifshitz solution (see for example the exact solution in appendix \ref{exactsolution} or the $AdS_4$ black hole/black brane solutions).  Again, for $z>2$, this cancels all infinite energy modes.  Hence, for $z>2$ we expect to obtain a one parameter family of finite energy solutions.

When $1\leq z \leq 2$ we again find that we must set both $\mc_1=0$ and $\mc_2=0$, constraining the two free parameters at the horizon.  Because of this, we expect to generically find at most a discrete set of solutions.  This would have been true even if we took arbitrary $\mc_1, \mc_2$ as the background solution: fixing $\mc_1=0, \mc_2=0$ is just a particular case of this. Let us consider looking for backgrounds with fixed values $\mc_{1,{\rm ref}}$ and $\mc_{2,{\rm ref}}$.  There are still two conditions at infinity and so we still need to tune both $g_0$ and $r_0$ so that both $\mc_1$ and $\mc_2$ match their reference values.  Therefore we would still expect at most a discrete set of $g_0,r_0$ that satisfy these conditions.  Other differences between the $z>2$ and $1<z<2$ were already found in \cite{Blau:2009gd}. The presence of the sphere in this case breaks the scaling symmetry.  The lack of a scaling argument does not allow us to make as strong a statement as for the $\sigma=0$ case. So, generically we expect a discrete set of solutions for $\sigma=1, \; 1<z<2$, analogous to the discrete set of Lifshitz stars found in \cite{Danielsson:2008gi}.

Given the above considerations, the program for finding black holes/branes is as follows.  We have two free parameters at the horizon, $g_0$ and $r_0$.  We consider fixing $r_0$ and scanning through possible values of $g_0$ until we find a solution that at infinity admits the expansion (\ref{A1expansion})-(\ref{G1expansion}).  We further scan through $g_0$ until the coefficient $\mc_1=0$ for our solutions, as we expect this mode to either give divergent behavior (for $z>2$) or have infinite energy (for any value of $z$).  This is sufficient to find finite energy (density) black holes/branes for the $z>2$ cases.  For the $1<z<2$ cases, we would also have to scan through $r_0$ until $\mc_2$ was zero as well.  However, we will be unable to do so, as we can only set $\mc_1$ numerically close to zero, and not actually zero.  We will not be sensitive to the power in front of $\mc_2$ because the non linear corrections to the linearized equations.

One could in principle do a higher order in perturbation theory calculation, and go to sufficiently high order that one could be sensitive to $\mc_2$.  We may get a ballpark estimate of how high in perturbation theory one would need to go before one would be numerically sensitive to these modes.  Let us take $z=3/2$ as an example.  In this case $-\frac{z}{2}-1+\frac{\gamma}{2}\approx-0.1492$ and $-\frac{z}{2}-1-\frac{\gamma}{2}\approx-3.3508$.  Generically, this term is present and we can only numerically set its coefficient $\mc_1$ to some small number.  Hence, to find out what the value of $\mc_2$ is, we would need to remove $r^{-\frac{z}{2}-1+\frac{\gamma}{2}}$ and all its subsequent descendent modes arising from nonlinearity in the system. These descendent modes will typically be integral powers of $r^{-\frac{z}{2}-1+\frac{\gamma}{2}}$, and so we ask the question, for what integer $n$ is $n\left(-\frac{z}{2}-1+\frac{\gamma}{2}\right)<\left(-\frac{z}{2}-1-\frac{\gamma}{2}\right)$ so that we are sensitive to the power multiplying $\mc_2$.  For the case of $z=3/2$, the answer is $n=23$, and so one would have to go to at least $22^{nd}$ order in perturbation theory separate out $\mc_2$!  For this reason, we do not attempt to do this here.  In fact closer to $z=1$ and $z=2$ this problem becomes worse because $-\frac{z}{2}-1+\frac{\gamma}{2}$ approaches $0$.

In what follows, we will simply set $\mc_1$ to zero, knowing this is sufficient for the $z>2$ cases and necessary for the $1<z<2$ cases.  We will also restrict to the $\sigma=1$ case: the $\sigma=0$ solutions can be obtained from the plots below by simply looking at large $r_0$: in this situation the term $e^{A(r)+C(r)}$ potential term always remains perturbative, and so is arbitrarily close to setting $\sigma=0$. Setting $\mc_1=0$ will furnish $g_0$ in terms of $r_0$ and so we can find the temperature via
\be
T=\frac{r_0^{z+1}a_0}{4\pi c_0}
\ee
where $a_0$ is chosen so that $e^{A_1(r)}$ asymptotes to $1$ rather than an arbitrary constant and $c_0$ is given in terms of $g_0$ and $r_0$ in (\ref{c0equ}).  Of course this is a ``unitless'' temperature: all units are restored with $L$.

Again, we stress that we do not have the numeric precision to be able to set the coefficient $\mc_2=0$ for the $1<z<2$ cases, as explained above.  Because of this extra constraint, we expect to get a discrete set of black holes.  Therefore, for $1<z<2$ the temperature $T(r_0)$ plots are to be regarded as the curve along which a discrete set of points represent good finite energy black holes; for $z>2$ the curves represent a one parameter family of black holes.
\begin{figure}[ht!]
\centering
    \includegraphics[width=0.45\textwidth,angle=0]{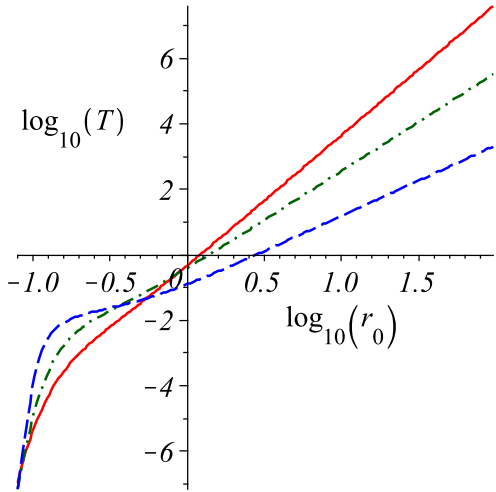}
\caption{ Graphs of $\log_{10}\left(T(r_0)\right)$ for $z=4,3,2.01$ colored  red (solid), green (dashdot) and blue (dashed) respectively for the $\sigma=1$ case.  Note the feature developing for smaller values of $z$.}
\label{TplotsZs}
\end{figure}

There is a possible exception to this: if for some reason setting $\mc_1=0$ also gives $\mc_2=0$ given the physical boundary conditions at the horizon, then these curves are actually the temperatures of a continuous set of black holes.  Although we feel that such a situation would be non generic, we leave it open as a possibility.  In figure \ref{TplotsZs} we plot the $T(r_0)$ for $z>2$ cases $z=4,3,2.01$.

\begin{figure}[ht!]
\centering
\subfloat[$z=1.1,z=1$]{\label{TplotsZ11}
    \includegraphics[width=0.3\textwidth,angle=0]{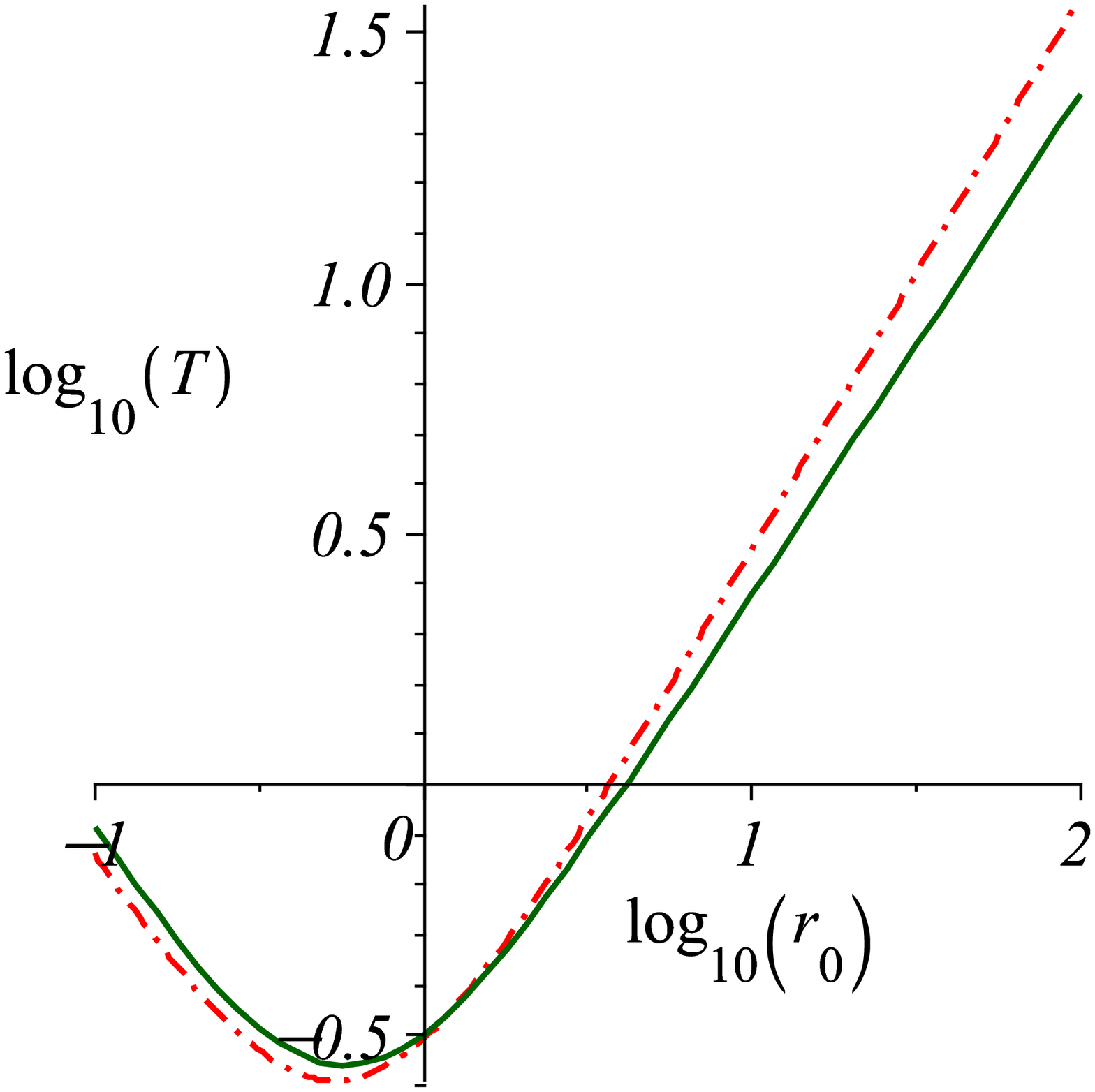}}
\subfloat[$z=1.74, 1.761, 1.78$]{\label{TplotsZcrit}
     \includegraphics[width=0.3\textwidth,angle=0]{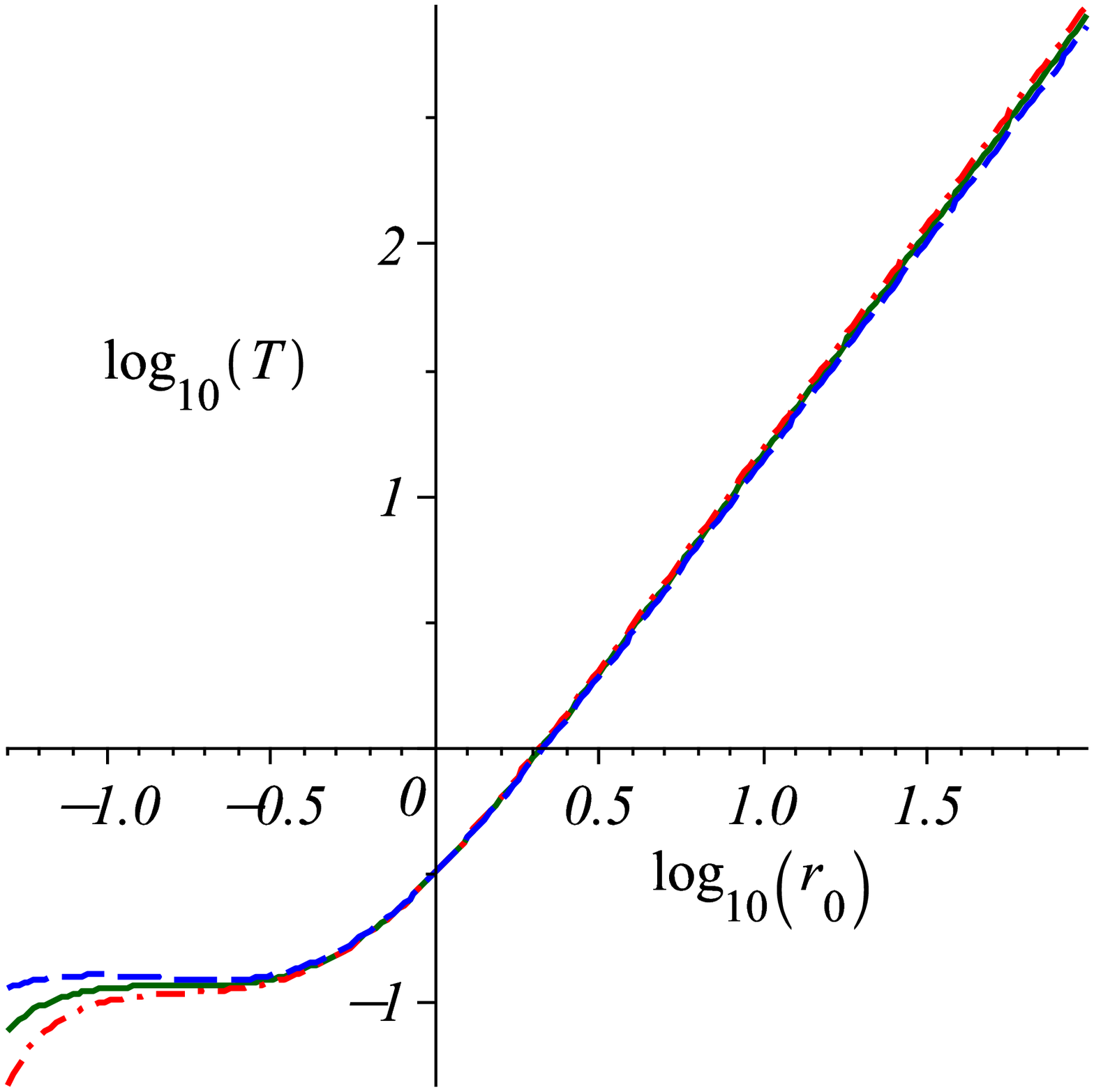}}
\caption{ On the left is graphs of $\log_{10}\left(T(r_0)\right)$ for $z=1.1$ colored red (dash-dot), and the temperature for $z=1$ the normal AdS black hole with temperature $T=\frac{3r_0^2+1}{4\pi r_0}$ in green (solid).  On the right, we plot $\log_{10}\left(T(r_0)\right)$ for $z=1.74, 1.761, 1.78$ in blue (dashed), green (solid), and red (dashdot) respectively.  All graphs are for the $\sigma=1$ case.}
\label{Tplotsflip}
\end{figure}

In figure \ref{TplotsZs} a flattening of the functions occurs at $\log_{10}(r_0)\sim-0.5$ for smaller $z$.  One may wonder whether these curves develop a negative slope, and so signal some sort of thermodynamic instability.  In figure \ref{TplotsZ11} we plot the $T(r_0)$ curve for $z=1.1$ and compare this to the pure AdS black hole, and indeed we find that near $\log_{10}(r_0)\sim -0.25$ the slope becomes negative.  If this represents a continuous set of black holes this would signal a thermodynamic instability: if however it is just the curve along which a discrete set of black holes may exist, then the interpretation is not obvious.

We also find the critical value of $z$ where a negative slope of $T(r_0)$ is first developed; we find this to be $z\approx 1.761$.  This is shown in figure \ref{TplotsZcrit}.  One further note is in order. Although $T(r_0)$ develops a negative slope for certain values of $r_0$, for sufficiently small $r_0$ it seems to become positive again.  If we take the negative slope as a sign of a thermodynamic instability, it does not have the runaway behavior of the pure AdS black holes.  It is eventually ``caught'' from shrinking further after it becomes sufficiently small.  This is, of course, assuming that we have a continuous set of black holes.

\begin{figure}[ht!]
\centering
\subfloat[$z=3,r_0=0.1$]{\label{smallZis3}
    \includegraphics[width=.3\textwidth,angle=0]{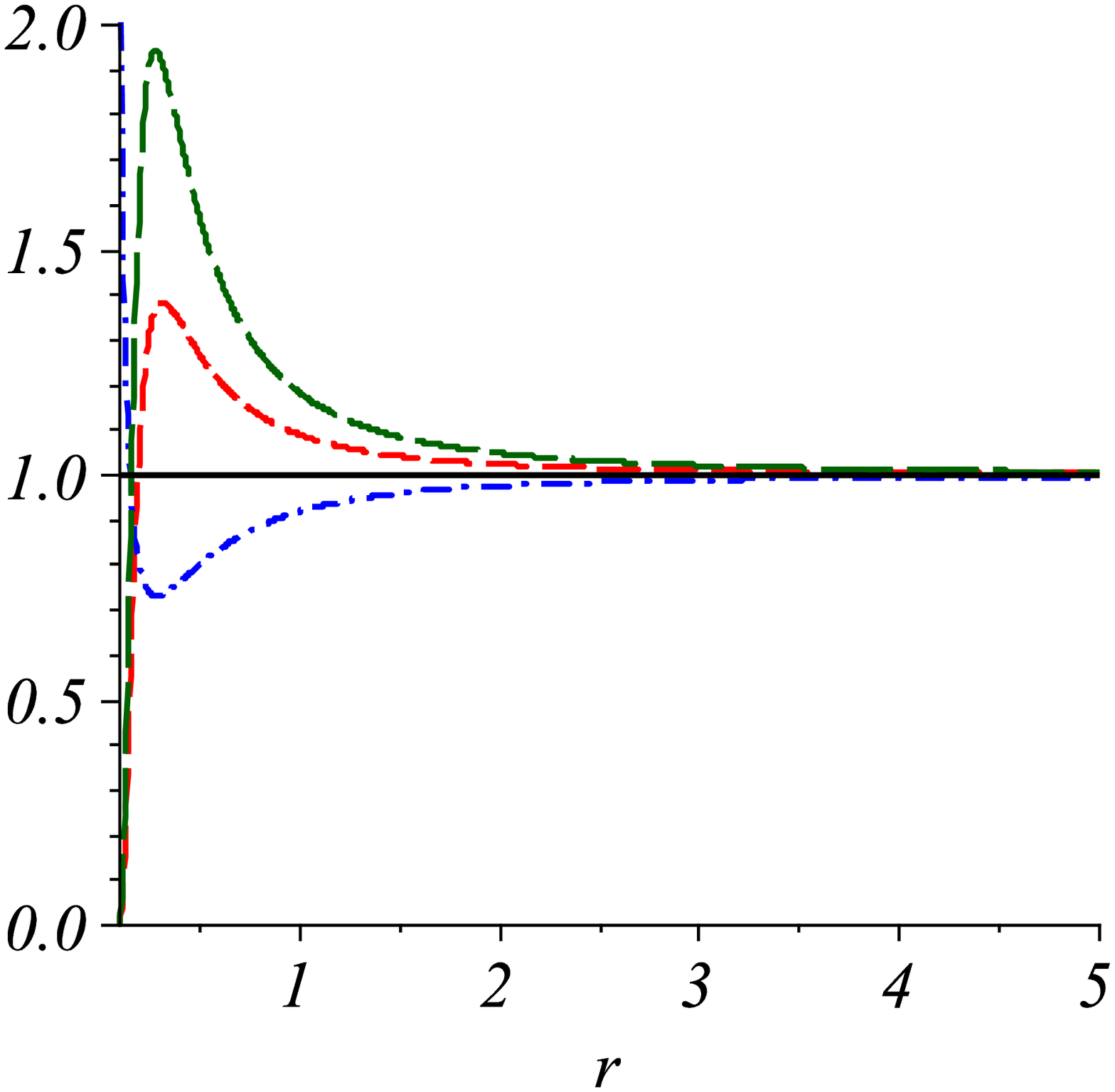}}
\subfloat[$z=3,r_0=5$]{\label{largeZis3}
    \includegraphics[width=.3\textwidth,angle=0]{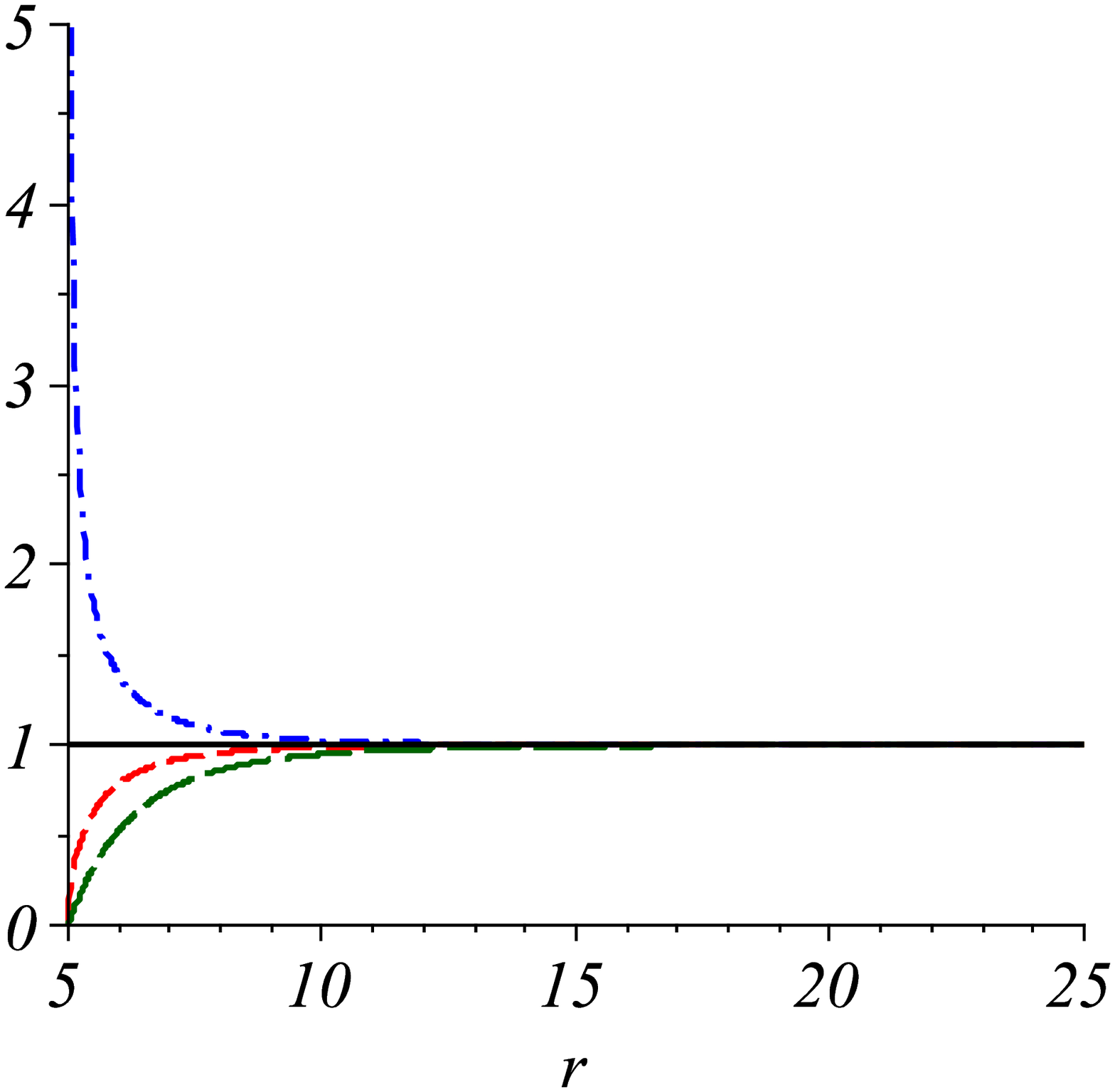}}\\
\subfloat[$r_0=0.1$, pure AdS]{\label{smallAdS}
    \includegraphics[width=.3\textwidth,angle=0]{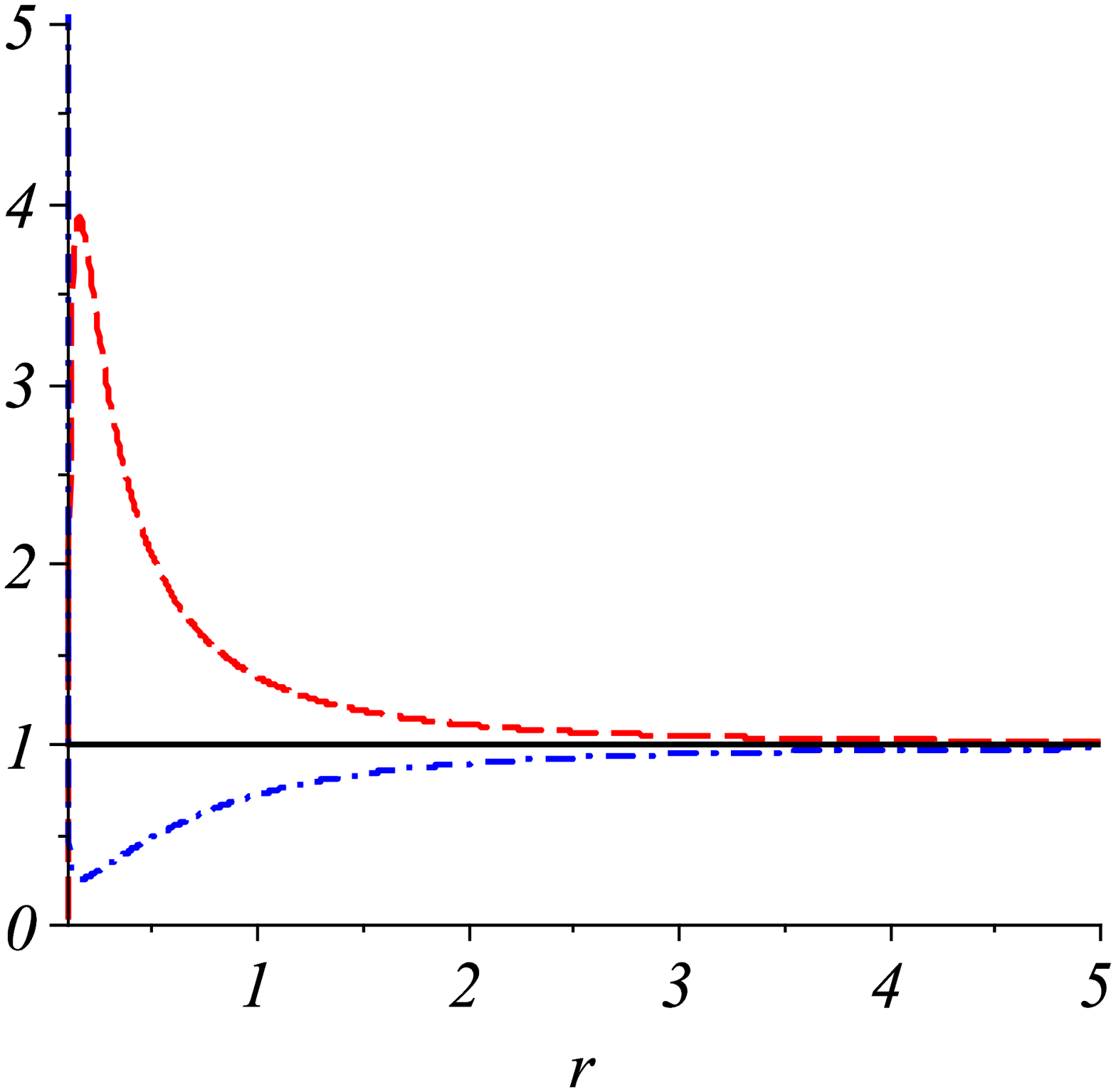}}
\subfloat[$r_0=5$, pure AdS]{\label{largeAdS}
    \includegraphics[width=.3\textwidth,angle=0]{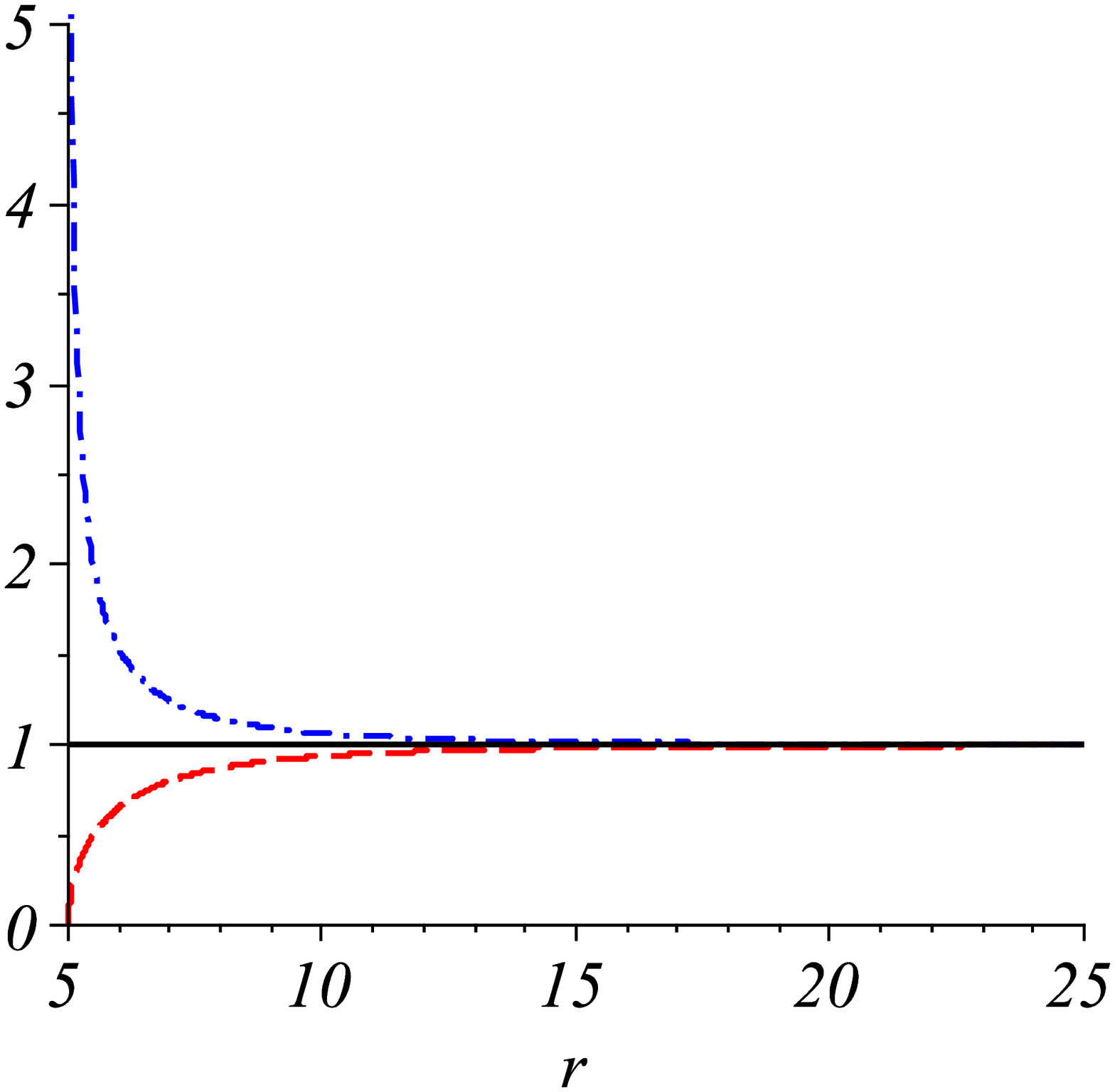}}
\caption{ Graphs of the black hole correction functions $e^{A_1(r)},e^{C_1(r)},e^{G_1(r)}$, colored red (short-dashed), blue (dash-dot) and green (long-dashed) respectively, for the $\sigma=1$ case.  The (solid) black flat line is to guide the eye to the asymptotic value of $1$ for all the functions.  Note that the large black holes remain monotonic, while the small black holes have a feature at several times $r_0$.  Further, it takes many more times $r_0$ for the solution to asymptote to $1$ in the ``small'' case.  This behavior is expected from the black hole in AdS, also graphed above for comparison.  The black brane $\sigma=0$ case is given by the large $r_0$ case above.}
\label{graphSmallLarge}
\end{figure}

For completeness, we also show several plots of the functions $e^{A_1(r)},e^{C_1(r)}$ and $e^{G_1(r)}$ in figure \ref{graphSmallLarge}.  One should notice the universal behavior for large $r_0$; the functions remain monotonic, and that for small $r_0$, that they develop an extra feature.  This is similar to the behaviors of small and large black holes in AdS, which are also graphed in figure \ref{graphSmallLarge}.
Again, the large $r_0$ behavior of all the graphs is expected to be identical with the ${\sigma}=0$ case, as the term in the potential $e^{A+C}$ always remains perturbative.  So we do not consider this a separate case: simply a limit of the solutions given here.

We conclude with some open problems.  The most striking is the question of whether for $1\leq z \leq 2$ finite energy black holes form a continuous family, or a discrete set.  This seems to be a difficult question to answer, even numerically, given the order in perturbation theory one would need to attain (discussed above).  Perhaps the most efficient way to address this problem is shooting in from finite energy conditions at infinity, and determining whether one develops a regular horizon.

One could also consider other ``finite energy'' conditions that one might be able to impose on the above solutions.  One possible candidate is the program of using local counter terms at the boundary \cite{Balasubramanian:1999re,Emparan:1999pm}, to remove the infinite energy contributions for black hole backgrounds \cite{Batrachenko:2004fd}.  Such a program might give some connection between the modes multiplied by $\mc_1$ and $\mc_2$.

Of further interest would be to embed the above action and/or solutions into a supergravity.  This could give some clue as to how these solutions are embedded into string theory, but also may open up new branches of solutions by introducing new matter content, and further one could look for possible supersymmetric versions of these backgrounds.  Particularly interesting is the construction of \cite{SchaferNameki:2009xr} where non relativistic spacetimes were realized as cosets of the Schr\"odinger algebra.  This raises the interesting question of whether these solutions can be constructed as the bosonic part of some supercoset models. We look forward to addressing these issues in the future.

\vspace*{0.5cm}
{{\bf note added:}\\
After releasing our work, the program of constructing local boundary
counter terms has been performed \cite{Ross:2009ar}. There, the
authors show that when the largest mode at infinity is set to zero
($\mc_1=0$) the energy is finite for any $z$.  This leads us to
conclude that the above black holes for $1<z<2$ are in fact a
continuous set.}
\vspace*{0.5cm}

\section*{Acknowledgements}

We wish to thank Peng Gao for his involvement at early stages of this work.  We also wish to thank Bob Holdom for useful discussions.  This work has been supported by NSERC of Canada.

\appendix

\section{An exact solution.}
\label{exactsolution}
In this section we will present an exact solution (we really mean an: there are no free parameters).  In fact, we are able to find 2 solutions, one with a naked singularity.  Since both of these solutions follow from similar guesswork, we will display them simultaneously.  If we start with equation (\ref{eCeAeq}) and require that $e^{C_1(r)}e^{A_1(r)}=1$ (or actually, any constant, but this may be removed with rescaling time), we in fact get the algebraic relation $e^{G_1}=\pm e^{2 A_1}$.  This we plug into (\ref{G1eq}) and get a {\it linear} equation in $e^{2A_1}$.  It reads
\be
-r^2\pa \pa e^{2A_1(r)}-r (z+3) \pa e^{2A_1(r)}-2ze^{2A_1(r)}+2z
\ee
and has the solution
\be
e^{2A_1(r)} = M_1r^{-z}+\frac{M_2}{r^2}+1
\ee
which converges to $1$ for large $r$.  We then plug this into either (\ref{A1eq}) or (\ref{C1eq}) and find
\be
-2 z+(4z-4z^2+2z^3)M_2+\frac{4r^{(4-z)}zM_1+(-5z^2+8z+z^3-4)M_2^2}{r^2}.
\ee
This can obviously not be solved for generic $z$.  However, for $z=2$ and for $z=4$ we do find solutions.
\be
e^{-2C_1(r)}=e^{G_1(r)}=e^{2A_1(r)}= \begin{cases} 1+\frac{1}{2r^2} & \text{if $z=2$} \\
                           1+\frac{1}{10r^2}-\frac{3}{400 r^4} & \text{if $z=4$} \end{cases}
\ee
Clearly the $z=4$ case has a horizon.  The $z=2$ case, however, does not (a recent paper \cite{Mann:2009yx} also has this solution).  One may be concerned about the fact that the two modes $1/r^2$ and $r^{-z}$ are the same in this case, and one must really resolve the equations and get a logarithmic piece. However, one can show that the coefficient to the logarithmic piece must be zero, and so one simply arrives at the $1/(2r^2)+1$ type of solution.  One can show that the $z=2$ solution has a curvature singularity at $r=0$ and that this at finite (spacelike) geodesic distance from any point in the space.  We therefore consider this an unphysical solution.  However, the $z=4$ solution is a good solution, with finite Ricci tensor, and finite Riemann square for all $r\geq r_0=\frac{1}{2\sqrt{5}}$, where this value of $r_0$ is the position of the horizon.  Both the Ricci tensor Riemann square tensors only become infinite at $r=0$.

For completeness, we write the metric and one-form explicitly here.
\bea
ds^2&&=-r^8\left(1+\frac{1}{10r^2}-\frac{3}{400 r^4}\right)dt^2+r^2d\Omega^2+\frac{dr^2}{r^2\left(1+\frac{1}{10r^2}-\frac{3}{400 r^4}\right)} \nn \\
\ma&&=Lr^4\sqrt{\frac{3}{2}}\left(1+\frac{1}{10r^2}-\frac{3}{400 r^4}\right)dt
\eea
where $d\Omega^2$ is the line element on the unit two sphere.

\section{Gauge invariance}
\label{gaugesection}
In the previous sections, we have gauge fixed by taking $B_1(r)=0$.  Here, we write down the linearized gauge transformations  that will allow us to switch to other gauges in perturbation theory (used near $r=\infty$).  The transformation
\bea
A_1(r)&&\rightarrow A_1(r)+\frac{z}{r}\delta(r)\nn \\
B_1(r)&&\rightarrow B_1(r)+\frac{1}{r}\delta(r)  \label{gaugesym}\\
C_1(r)&&\rightarrow C_1(r)-\frac{1}{r}\delta(r)+\pa_r \delta(r)\nn \\
G_1(r)&&\rightarrow G_1(r)+\frac{z}{r}\delta(r) \nn
\eea
corresponds to infinitesimal coordinate transformations $r\rightarrow r+ \epsilon \delta(r)$.  One can see that such a shift leaves the first order equations (near $r=\infty$) unchanged by explicitly plugging in the above shift.

\end{document}